\documentclass[aps,prl,twocolumn,showpacs,preprintnumbers,amsmath,amssymb]{revtex4}   

\usepackage{graphicx}
\usepackage{dcolumn}
\usepackage{bm}

\newcommand {\beq} {\begin{equation}}
\newcommand {\eeq} {\end{equation}}
\newcommand {\beqa}{\begin{eqnarray}}
\newcommand {\eeqa}{\end{eqnarray}}
\newcommand {\del} {\partial}
\newcommand {\tr}{{\rm tr\,}}

\newcommand {\ee}{\mbox{e}}

\begin{document}


\title{
Monte Carlo studies
of supersymmetric matrix quantum mechanics\\
with sixteen supercharges at finite temperature}
 
\author{Konstantinos N. Anagnostopoulos$^{1}$}
\email{konstant@mail.ntua.gr}
\author{Masanori Hanada$^{2}$}
\email{hana@riken.jp}
\author{Jun Nishimura$^{3,4}$}
\email{jnishi@post.kek.jp}
\author{Shingo Takeuchi$^{4}$}
\email{shingo@post.kek.jp}

\affiliation{
$^{1}$Physics Department,  National Technical University of Athens,
Zografou Campus, GR-15780 Athens, Greece \\
$^{2}$Theoretical Physics Laboratory, 
RIKEN Nishina Center,
2-1 Hirosawa, Wako, Saitama 351-0198, Japan \\
$^{3}$High Energy Accelerator Research Organization (KEK), 
		Tsukuba 305-0801, Japan \\
$^{4}$Department of Particle and Nuclear Physics,
School of High Energy Accelerator Science,
Graduate University for Advanced Studies (SOKENDAI),
Tsukuba 305-0801, Japan
}

\date{July, 2007; preprint: RIKEN-TH-112, KEK-TH-1165
}


\begin{abstract}
We present the first Monte Carlo results
for supersymmetric matrix quantum mechanics with
sixteen supercharges at finite temperature.
The recently proposed {\em non-lattice} simulation
enables us 
to include the effects of fermionic matrices
%
%
in a transparent and reliable manner.
%
%
The internal energy nicely interpolates the
weak coupling behavior obtained by the high temperature
expansion, and the strong coupling behavior 
predicted from the dual black hole geometry.
The Polyakov line takes large values even at low temperature
suggesting the absence of a phase transition 
in sharp contrast to the bosonic case.
These results provide highly non-trivial evidences 
for the gauge/gravity duality.
%
\end{abstract}


\maketitle


\paragraph*{Introduction.---}

In the last decade, we have witnessed
the increasing
importance of large-$N$ gauge theories
in 
theoretical particle physics.
For instance,
the 
holographic principle, which was inspired
originally by the Bekenstein-Hawking
formula for the black hole entropy,
has now been given a concrete manifestation 
as a conjectured 
duality between the strongly coupled
large-$N$ gauge theory and
the weakly coupled supergravity.
The best understood example is the
AdS/CFT correspondence \cite{Maldacena:1997re},
but there are numerous extensions
to non-conformal field theories as well.
In particular, large-$N$ gauge theories
in low dimensions have been studied
intensively at finite temperature,
which revealed
intriguing connections
to the black hole thermodynamics 
\cite{Itzhaki:1998dd,Barbon:1998cr,%
KLL,Aharony:2004ig,Aharony4}.

Large-$N$ gauge theories 
in low dimensions
also play an important role
in formulating superstring/M theories 
non-perturbatively
based on the idea of matrix models,
which was successful 
in the case of non-critical strings.
%
For instance, it is conjectured that
critical string/M theories
can be formulated in terms of matrix models, 
which can be formally obtained
by dimensionally reducing 
U($N$) super Yang-Mills theory 
in ten dimensions
to $D=0,1,2$ dimensions.
%
In particular, the $D=1$ case \cite{BFSS}
corresponds to
the M Theory \cite{Witten:1995ex},
which is a hypothetical eleven-dimensional
theory proposed to understand the dualities
among all the known superstring theories 
in ten dimensions.

In order to confirm these conjectures
or to make use of them,
it is clearly important to
study large-$N$ gauge theories from 
first principles.
Monte Carlo simulation is expected to
be very useful for that purpose.
In particular, totally reduced
models \cite{IKKT} (the $D=0$ case) 
have been studied in refs.\ 
\cite{KNS,HNT,Ambjorn:2000dx,%
Anagnostopoulos:2001yb,Burda:2000mn}.
The complex Pfaffian, which appears from integration
over the fermionic matrices,
causes a technical obstacle
in numerical simulation,
which may be overcome by a new method 
proposed in ref.\ \cite{Anagnostopoulos:2001yb}.
In fact, the phase of the Pfaffian is speculated to
induce the spontaneous breaking of SO(10)
symmetry down to SO(4), a scenario for the
dynamical generation of four-dimensional
space-time \cite{AIKKT} suggested 
from the Gaussian expansion method \cite{SSB}.

In the $D=1$ case, some sort of 
``discretization''
is needed in order to put the theory
on computer. Given the well-known problems
with the conventional lattice discretization,
three of the authors 
(M.H., J.N.\ and S.T.)
have proposed a {\em non-lattice}
simulation method \cite{Hanada-Nishimura-Takeuchi}, 
which is useful 
for studying supersymmetric quantum mechanics.
The crucial point was that the gauge symmetry
is almost trivial in 1d, and therefore, we
can fix the gauge completely.
This allows us to introduce a Fourier mode cutoff 
$\Lambda$ without violating the gauge symmetry.
In the bosonic case 
the new method reproduced the lattice results
in the continuum limit.
In the SUSY case with four supercharges,
it reproduced
the results of the high temperature expansion
in the continuum.
The same model has been studied 
in ref.\ \cite{Catterall:2007fp}
by the lattice approach using a simple
lattice action and a more complicated 
lattice action preserving half of SUSY. 

In this work we apply the non-lattice
simulation method to the most interesting case
with sixteen supercharges.
%
We have reduced the computational effort
considerably
by introducing pseudo-fermions
based on the idea of 
the Rational Hybrid Monte
Carlo (RHMC) algorithm \cite{Clark:2004cp}
in the way described 
in ref.\ \cite{Catterall:2007fp}.

As discussed in ref.\
\cite{Hanada-Nishimura-Takeuchi},
our action is nothing but the
gauge-fixed action in the continuum
except for having a Fourier mode cutoff.
Supersymmetry, which is mildly broken by
the cutoff, is shown to be restored
much faster than the continuum limit is
achieved.
In fact, the continuum limit is also
approached faster than one would naively expect
from the number of degrees of freedom.
This is understandable from the fact that
the modes above the cutoff are naturally
suppressed by the kinetic term.
A further (albeit technical) 
advantage of our formulation
is that we can
implement the Fourier acceleration,
which eliminates the 
critical slowing down completely 
\cite{Catterall:2001jg},
{\em without extra cost}
since we are dealing with 
Fourier modes directly.
We consider that all the theoretical
and technical merits of the present
approach compensate the superficial
increase in the computational effort 
by the factor of O($\Lambda$)
compared to the lattice approach
with the same number of degrees of freedom.

\paragraph*{Simulation techniques.---}

The model 
can be obtained formally by dimensionally
reducing 10d super Yang-Mills theory to 1d.
The action is given by
\beqa
S
&=& 
\frac{1}{g^2} \int_0^{\beta}  
d t \, 
\tr 
\bigg\{ 
\frac{1}{2} (D_t X_i)^2 - 
\frac{1}{4} [X_i , X_j]^2  
\nonumber \\
&~& 
+ \frac{1}{2} \psi_\alpha D_t \psi_\alpha
- \frac{1}{2} \psi_\alpha (\gamma_i)_{\alpha\beta} 
 [X_i , \psi_\beta ]
\bigg\} \ ,
\label{cQM}
\eeqa
where $D_t  = \del_t
  - i \, [A(t), \ \cdot \ ]$ represents the covariant derivative
with the gauge field $A(t)$ being an $N\times N$ Hermitian matrix.
It can be viewed as 
a one-dimensional U($N$) gauge theory with adjoint matters.
The bosonic matrices $X_i(t)$  $(i=1,\cdots,9)$
come from spatial components of the 10d gauge field,
while the fermionic matrices $\psi_\alpha(t)$
$(\alpha=1,\cdots , 16)$ come from
a Majorana-Weyl spinor in 10d.
The $16\times 16$ matrices $\gamma_\mu$
in (\ref{cQM}) act on spinor indices and
satisfies the Euclidean Clifford algebra
$\{ \gamma_i,\gamma_j \}= 2\delta_{ij}$.
We impose periodic and anti-periodic
boundary conditions
on the bosons and fermions, respectively.
The extent $\beta$ in the Euclidean time 
direction then corresponds to the inverse
temperature $\beta \equiv 1/T$.
The parameter $g$ in (\ref{cQM})
can always be scaled out by
an appropriate rescaling of the matrices and 
the time coordinate $t$.
We take $g = \frac{1}{\sqrt{N}}$ without loss 
of generality.


We take the static diagonal gauge
$A(t) = \frac{1}{\beta} {\rm diag}
(\alpha_1 , \cdots , \alpha_N)$,
where $\alpha_a$ can be chosen to 
satisfy
the constraint 
$\max_a (\alpha_a) - \min_a (\alpha_a) 
\le 2\pi$
by using the large gauge transformation
with a non-zero winding number.
We have to add to the action a term
\beq
S_{\rm FP} =
- \sum_{a<b} 2 \ln 
\left| \sin \frac{\alpha_a - \alpha_b}{2}
\right|  \ , 
\eeq
which appears from the Faddeev-Popov procedure,
and the integration measure for $\alpha_a$
is taken to be uniform.

We make a Fourier expansion 
\beq
X_i ^{ab} (t) = \sum_{n=-\Lambda}^{\Lambda} 
\tilde{X}_{i n}^{ab} \ee^{i \omega n t} \ ; \
\psi_\alpha ^{ab} (t) = \sum_{r=-\lambda}^{\lambda}
\tilde{\psi}_{\alpha r}^{ab} \ee^{i \omega r t} \ ,
\eeq
where $\lambda \equiv \Lambda-1/2$.
The indices $n$ and $r$ take integer and
half-integer values, respectively,
corresponding to the imposed 
boundary conditions.
Introducing a shorthand
notation
\beq
\Bigl(f^{(1)}  \cdots  f^{(p)}\Bigr)_n 
\equiv \sum_{k_1 + \cdots + k_{p}=n}
f^{(1)}_{k_1} \cdots f^{(p)}_{k_p} \ ,
\eeq
we can write the action 
(\ref{cQM}) as
$S=S_{\rm b}+S_{\rm f}$, where
\beqa
S_{\rm b}
&=&  N \beta
\Bigg[
\frac{1}{2} \sum_{n=-\Lambda}^{\Lambda} 
\left( n \omega - \frac{\alpha_a - \alpha_b}{\beta} 
\right)
^2   \tilde{X}_{i , -n}^{ba} \tilde{X}_{i n}^{ab}
\nonumber \\ 
&~& 
- \frac{1}{4} 
 \tr \Bigl( [ \tilde{X}_{i} , \tilde{X}_{j}]^2  \Bigr)_0
\Bigg] 
\nonumber \\
S_{\rm f}
&=& \frac{1}{2}
N \beta \sum_{r=-\lambda}^{\lambda} \Biggl[
i 
\left(
r \omega - \frac{\alpha_a - \alpha_b}{\beta} 
\right)
\tilde{\bar{\psi}}_{\alpha r}^{ba} \tilde{\psi}_{\alpha r}^{ab} 
 \nonumber \\
&~& - (\gamma_i)_{\alpha\beta}
 \tr \Bigl\{ \tilde{\bar{\psi}}_{\alpha r} \Bigl(
[ \tilde{X}_{i},\tilde{\psi}_{\beta}] \Bigr)_r \Bigr\} \Biggr] \ .
\label{bfss_action_cutoff}
\eeqa
%
The fermionic action $S_{\rm f}$ may be
written in the form
$S_{\rm f}
= \frac{1}{2} {\cal M}_{A \alpha r ; B \beta s}
\tilde{\psi}_{\alpha r}^A \tilde{\psi}_{\beta s}^B
$,
where we have expanded
$\tilde{\psi}_{\alpha r}
= \sum_{A=1}^{N^2} \tilde{\psi}_{\alpha r}^A t^A$
in terms of U($N$) generators $t^A$.
Integrating out the fermionic variables, one obtains
the Pfaffian ${\rm Pf}{\cal M}$, which is complex
in general. However, we observe that it is actually
real positive with high accuracy in the temperature
regime studied in the present work.
Hence we can replace it by 
$|{\rm Pf}{\cal M}|
= {\rm det} ( {\cal D}^{1/4})$,
where ${\cal D}={\cal M}^\dag {\cal M}$.

The trick of the RHMC algorithm is to
use the rational approximation
$x^{-1/4} \simeq
b_0 + \sum_{k=1}^{Q}
\frac{a_k}{x+b_k} 
$,
which has sufficiently small relative error 
within a certain range required by the system
to be simulated.
%
(The real positive parameters 
$a_k$ and $b_k$ can be
obtained by a code \cite{Clark-Kennedy} based on 
the Remez algorithm.)
Then the Pfaffian is replaced by
$
|{\rm Pf}{\cal M}|
= \int dF dF^* \ee^{-S_{\rm PF}} 
$, where
\beq
S_{\rm PF} =
b_0 F^* F + \sum_{k=1}^{Q}
a_k F^* ({\cal D}+b_k)^{-1} F \ ,
\label{PF-pf}
\eeq
using the auxiliary complex variables $F$,
which is called the 
pseudo-fermions.

We apply the usual
HMC algorithm to the whole system
as described in ref.\
\cite{Hanada-Nishimura-Takeuchi},
except that now we introduce
the momentum variables conjugate to
the pseudo-fermions $F$ as well 
as the bosonic matrices $\tilde{X}_i$ 
and the gauge variables $\alpha_a$.
When we solve the auxiliary classical
Hamiltonian dynamics,
it is important to 
apply the Fourier acceleration 
\cite{Catterall:2001jg}
to the pseudo-fermions $F$ 
and the bosonic matrices $\tilde{X}_i$.
The main part of the computation
comes from
solving a linear system
$({\cal D}+b_k) \chi = F \quad
(k=1, \cdots , Q )
$.
We solve the system
for the smallest $a_k$
using the conjugate gradient
method, which reduces the problem to
the iterative multiplications of ${\cal M}$
to a pseudo-fermion field,
each of which requires O($\Lambda^2 N^3$)
arithmetic operations if implemented carefully.
The solution for larger $b_k$'s 
can be obtained as by-products
using the idea of the multi-mass 
Krylov solver \cite{Jegerlehner:1996pm}. 
This avoids the factor
of $Q$ increase of the computational effort.
%

%

\paragraph*{Infrared instability.---}

Since the integration domain for the
bosonic matrices is non-compact,
the convergence of the partition function
is not obvious.
In particular, there exists a potential
danger in the flat direction
corresponding to commuting matrices.
Such an issue has been addressed
in the totally reduced model 
\cite{AIKKT,KNS,HNT,AW}.
In the present $D=1$ case, 
let us expand the
cutoff theory
(\ref{bfss_action_cutoff})
around the commuting background
$
\tilde{X}_{i 0}
= {\rm diag} (
x_{i 1} , \cdots , x_{i N}
)
$
and consider the effective action 
for the moduli parameters $x_{i a}$
and $\alpha_a$. 
When both $T$ and all of 
$|x_{i a} - x_{i b}|$ are large,
the fluctuations become very massive,
and the one-loop approximation
is justified.
We can easily obtain
\beq
W_{\rm 1-loop} = 
 \sum_{a<b} 4 \log \left(
\frac{
\prod_n \{ (2 \pi n - \alpha_{ab} )^2
+ (\beta x_{ab})^2 \} 
}{
\prod_r \{ (2 \pi r - \alpha_{ab} )^2
+ (\beta x_{ab})^2 \} 
}
\right)  \ ,
\label{1-loop-eff}
\eeq
where we have defined
$\alpha_{ab} \equiv \alpha_a - \alpha_b$
and 
$x_{ab} \equiv \sqrt{(x_{ia} - x_{ib})^2}$.
In eq.\ (\ref{1-loop-eff})
we have omitted terms independent of $x_{ab}$,
which actually
vanish in the $\Lambda \rightarrow \infty$ 
limit.
When $x_{ab} \ll T$ and $\alpha_{ab} \ll 2 \pi $,
the $n=0$ term dominates and yields
a logarithmic attractive potential
$W_{\rm 1-loop} \simeq
\sum_{a<b} 4 \log 
\{ (\alpha_{ab} )^2
+ (\beta x_{ab})^2 \}  
$
among $\alpha_a$ and among $x_{ia}$.
This agrees with the well-known result
in the bosonic IKKT model, which describes
the high temperature limit of the present
model.
In fact one obtains $x_{ab} \sim T^{1/4}$
according to the high temperature 
expansion (HTE) \cite{HTE}.
On the other hand, when
$T \ll x_{ab} \ll 2 \pi \Lambda T $,
the denominator and the numerator
in eq.\ (\ref{1-loop-eff})
cancel each other almost completely.
This implies the existence of 
an instability.

As $T$ is lowered, the instability 
region approaches the peak at
$x_{ab} \sim T^{1/4}$
representing the high temperature behavior.
However, since we have a sum over all the
pairs of indices $(a,b)$, a tiny difference
between the denominator and the numerator is
enhanced by the factor of $N^2$.
Then it follows that
the lower edge of the instability region
gets
multiplied by $N$.
We found empirically that
the instability can be avoided 
by taking $N \gtrsim \frac{6}{T}$, which is
consistent with the above considerations.

At low temperature, one also has to worry
about the finite $\Lambda$ effects.
In the case of energy, 
they are negligible for 
$\Lambda \gtrsim \frac{2}{T}$, whereas
for the other observables 
studied in this paper,
we need twice as large $\Lambda$.



\paragraph*{Results.---}

\begin{figure}[htb]
\begin{center}
\includegraphics[height=6cm]{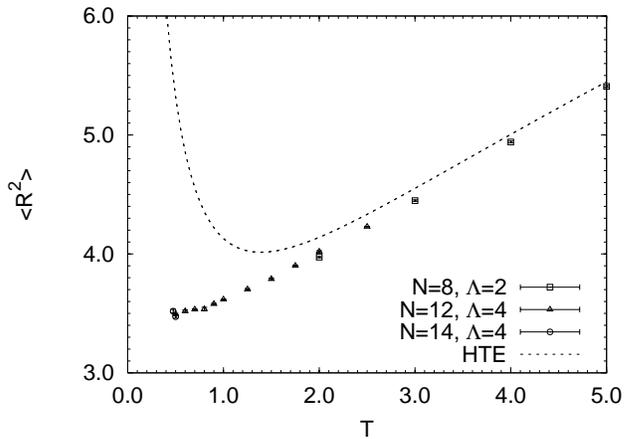}
\end{center}
\caption{
The ``extent of space'' is plotted 
against temperature.
The dashed line represents the result
obtained by HTE
up to the next leading order for $N=8$ \cite{HTE}.
}
\label{trX2}
\end{figure}

Fig.\ \ref{trX2} shows the average of
the ``extent of space''
$R^2 = 
\frac{1}{N \beta}
\int dt \, \tr (X_i)^2 
$.
The instability mentioned above
can be probed by the divergence
of this quantity.
We obtain stable results at
sufficiently large $N$.


\begin{figure}[htb]
\begin{center}
\includegraphics[height=6cm]{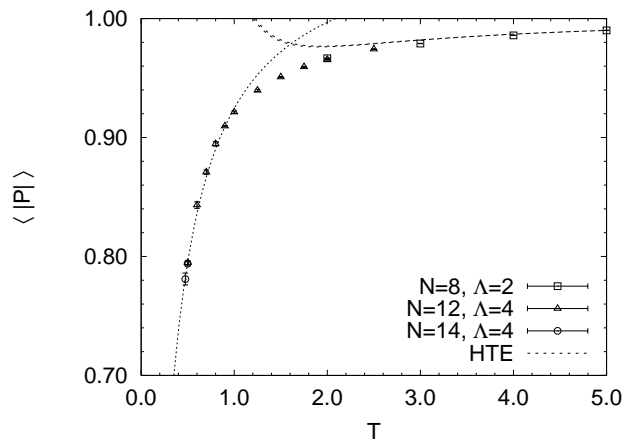}
\end{center}
\caption{
The absolute value of
the Polyakov line
is plotted against 
temperature.
The dashed line represents the result
obtained by HTE
up to the next leading order for $N=8$ \cite{HTE}.
The dotted line represents a fit to
eq.\ (\ref{polya-fit}) with $a=0.15$ and $b=0.072$.
%
%
}
\label{polyakov}
\end{figure}

In fig.\ \ref{polyakov}
we plot the absolute value 
of the Polyakov line, which is the
order parameter for the SSB of
the U(1) symmetry.
Unlike in the bosonic case 
\cite{Janik:2000tq,Aharony:2004ig,Kawahara:2007fn}, 
where the SSB occurs around $T \simeq 1$,
the Polyakov line is not small
even at low $T$.
This implies that there is no phase transition
in the SUSY case,
as predicted by the gauge/gravity 
correspondence \cite{Barbon:1998cr,Aharony4}.
We find that the Polyakov line can be fitted
nicely to
\beq
\langle |P| \rangle = \exp(-\frac{a}{T}+b) \ ,
\label{polya-fit}
\eeq
a characteristic behavior in
a deconfined theory.

\begin{figure}[htb]
\begin{center}
\includegraphics[height=6cm]{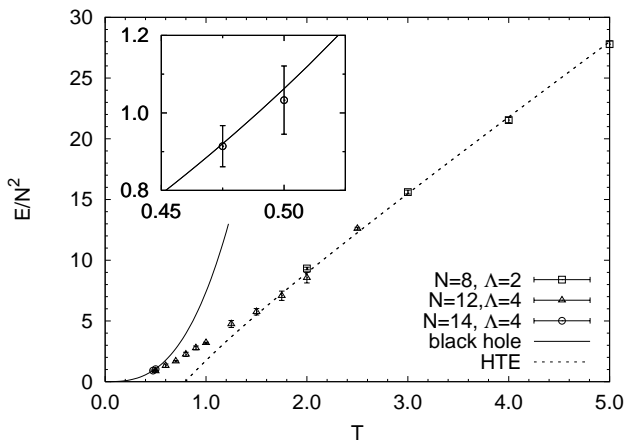}
\end{center}
\caption{
The energy
is plotted against 
temperature.
The dashed line represents the result
obtained by HTE
up to the next leading order 
for $N=8$ \cite{HTE}.
The solid line represents
the energy predicted at small $T$
by the gauge/gravity duality.
The upper left panel zooms
up the region, where
the power-law behavior sets in.
%
}
\label{energy}
\end{figure}

In fig.\ \ref{energy}
we plot the internal energy defined by
$E = \frac{\del}{\del \beta } (\beta {\cal F})$,
where ${\cal F}$ is the free energy
of the system.
In practice, we calculate it using a
formula, which follows from a simple scaling 
argument \cite{Catterall:2007fp}.
In our case it reads 
$E = 
- 3 T \left[ \langle S_{\rm b} \rangle
- \frac{9}{2} \{  (2 \Lambda +1) N^2 -1 \} 
\right]$.
Our results interpolate nicely the
weak coupling behavior --- calculated
by the HTE up to the
next leading order \cite{HTE} ---
and the strong coupling behavior 
$E/N^2
= 7.4 \cdot T^{2.8}$
predicted by the gauge/gravity duality 
\cite{Itzhaki:1998dd} from
the dual black-hole geometry \cite{Klebanov:1996un}.
The power-law behavior sets in
at $T \simeq 0.5$, 
which is reasonable since
the effective coupling constant
is given by
$\lambda_{\rm eff}=1/T^3$ in our convention.

In ref.\ \cite{KLL}
the Gaussian expansion method 
was applied to the present model,
and the energy 
obtained at the leading order
was fitted nicely to the power law
$E/N^2 = 3.2 \cdot T^{2.7}$ within
$0.25 \lesssim T \lesssim 1$.
Their results are in reasonable agreement
with our data at $T \sim 1$, but
disagree at lower temperature.


\paragraph*{Summary.---}

In this paper we have presented
the first Monte Carlo results
for the maximally supersymmetric matrix
quantum mechanics, which is expected to
play a very important role in string/M theories.
The recently proposed non-lattice simulation
together with the RHMC algorithm
enabled us to study the low temperature 
behavior, which was not
accessible by the high temperature expansion.
As we lower the temperature,
we observed the infrared instability,
which was found to be eliminated, however,
by increasing $N$. We gave a natural explanation
to this phenomenon 
based on the one-loop effective action.
Our data for the internal energy 
asymptote nicely to the result
obtained from the dual geometry, which we consider
as a highly nontrivial evidence for the
gauge/gravity duality
in the non-conformal case.
In particular, our results 
suggest that the maximally supersymmetric 
matrix quantum mechanics
exactly reproduces not only the power but also 
the coefficient
of the power-law behavior 
obtained from the dual black-hole geometry.
%





\paragraph*{Acknowledgments.---}
The authors would like to thank 
Shoji Hashimoto and Hideo Matsufuru
for helpful suggestions concerning
the RHMC simulation.
The computations were carried out on
supercomputers
(SR11000 at KEK,
SX8 at RCNP and SX7 at RIKEN)
as well as on PC clusters.
This work is supported by the EPEAEK programmes 
``Pythagoras II'' and co-funded
by the European Union (75\%) and 
the Hellenic state (25\%).



\end{document}